\documentclass{aa}

\usepackage{txfonts}
\usepackage{graphicx}
\usepackage{natbib}
\usepackage{amssymb}

\bibpunct{(}{)}{;}{a}{}{,} 

\begin{document}

\title        {The most metal-poor damped Ly $\alpha$ system at $z<3$: constraints
               on early nucleosynthesis}

\subtitle {}
\author   {P. Erni,     \inst{1}
           P. Richter,  \inst{1}
           C. Ledoux,   \inst{2} \and
	   P. Petitjean \inst{3,4}
	  }

\offprints {P. Erni\\
  \email{perni@astro.uni-bonn.de}}

\institute{Argelander-Institut f\"ur Astronomie 
           \thanks{Founded by merging of the Institut f\"ur Astrophysik und 
	   Extraterrestrische Forschung, the Sternwarte, and the Radioastronomisches 
	   Institut der Universit\"at Bonn.}, Universit\"at Bonn,
           Auf dem H\"ugel 71, 53121 Bonn, Germany
     \and  European Southern Observatory, Alonso de C\'{o}rdova 3107, Casilla
           19001, Vitacura, Santiago, Chile
     \and  Institut d'Astrophysique de Paris - CNRS, 98bis Boulevard Arago,
           75014, Paris, France
     \and  LERMA, Observatoire de Paris-Meudon, 61 avenue de l'Observatoire, 
           75014 Paris, France}

\date{Received 10 October 2005 / Accepted 27 January 2006 }

\abstract{
To constrain the conditions for very early nucleosynthesis in the Universe we compare
the chemical enrichment pattern of an extremely metal-poor damped Lyman $\alpha$ (DLA) 
absorber with predictions from recent explosive nucleosynthesis model calculations. For this, 
we have analyzed chemical abundances in the DLA system at $z_{\mathrm{abs}}=2.6183$ toward the quasar 
Q0913+072 ($z_{\mathrm{em}}=2.785$) using public UVES/VLT high spectral resolution data. The 
total neutral hydrogen column density in this absorber is log\,$N$(H\,{\sc i})\,=\,$20.36\pm0.05$.
Accurate column densities are derived for C\,{\sc ii}, N\,{\sc i}, O\,{\sc i}, Al\,{\sc ii},
Si\,{\sc ii}, and Fe\,{\sc ii}. Upper limits are given for Fe\,{\sc iii} and 
Ni\,{\sc ii}. With $[\rm{C/H}]=-2.83\pm0.05$, $[\rm{N/H}]=-3.84\pm0.11$, and $[\rm{O/H}]=-2.47\pm0.05$, 
this system represents one of the most metal-poor DLA systems investigated so far.
It offers the unique opportunity to measure accurate CNO abundances in a protogalactic 
structure at high redshift. Given the very low overall abundance level and the observed abundance 
pattern, the data suggest that the chemical evolution of this DLA system is dominated by
one or at most a few stellar generations.  With reference to numerical model
calculations, the chemical abundances in the DLA system are consistent 
with an enrichment from a single starburst of a zero-metallicity population of
massive stars ($\sim 10$-$50\,M_\odot$) exploding as core-collapse Supernovae (SNe), i.e., the
classical Type II Supernovae (SNe~II), and possibly as hyper-energetic 
($E>10^{\mathrm{51}}$\,erg) core-collapse Supernovae, so-called Hypernovae (HNe), as well. 
In contrast, models using non-zero metallicity progenitors or other explosion mechanisms, 
such as pair-instability Supernovae (PISNe) or Type Ia Supernovae (SNe~Ia), do not match the 
observed abundance pattern. Comparing our results with recent estimates for the global chemical 
evolution of the intergalactic medium (IGM) and early galactic structures  
shows that the observed metal abundances in the DLA system toward Q0913+072 are only slightly above the level
expected for the intergalactic medium (IGM) at $z\approx2.6$, but significantly lower than what 
is expected for the interstellar medium (ISM) in galaxies at that redshift. This
implies that this DLA system has recently condensed out of the IGM and
that local star formation in this protogalaxy has not yet contributed significantly
to the metal budget in the gas.

}

\titlerunning {Constraints on early nucleosynthesis}

\maketitle
\section{Introduction}

Recent theoretical studies (e.g., Omukai \& Palla 2003) predict that
the first (Pop~III) stars must have been very massive 
($\sim100$-$600\,M_\odot$), thus indicating an initial mass function (IMF)
that favors the formation of more massive stars (top-heavy IMF). Stellar evolution 
studies (Heger \& Woosley 2002) show that primordial stars with main-sequence (ms) 
masses between  $\sim 50$-$140\,M_\odot$ and above $\sim260\,M_\odot$ inevitably collapse into 
black holes and are unable to eject their metals. Furthermore, stars with masses
below $M_{\mathrm{ms}}\sim 10\,M_\odot$ do not significantly contribute to
the chemical feedback on galactic scales (Ciardi \& Ferrara 2005). Hence, 
only massive stars of $\sim 10$-$50\,M_\odot$, exploding as core-collapse S/HNe, 
or super-massive stars of $\sim 140$-$260\,M_\odot$, exploding as PISNe, will 
eventually enrich the ISM and subsequently the IGM. 

It has been proposed that metal enrichment is the mechanism responsible for a 
transition from a top-heavy to a more conventional power-law IMF as observed 
in the present-day Universe (Bromm et al. 2001; Schneider et al. 2002).
The metallicity is believed to reach a critical value at $[\rm{Z_{cr}/H}]\approx-4$, 
marking a transition from a high-mass to a low-mass fragmentation mode of the 
protostellar gas cloud. Conversely, recent observations of hyper metal-poor (HMP) stars, 
e.g., HE0107$-$5240 with $[\rm{Fe/H}]=-5.2\pm0.02$ (Christlieb et al. 2002) or 
HE1327$-$2326 with $[\rm{Fe/H}]=-5.4\pm0.02$ (Frebel et al. 2005) might, 
at first sight, rule out a top-heavy IMF for Pop~III stars.  
Other scenarios like a bimodal IMF (Nakamura \& Umemura 2001; 
Omukai \& Yoshii 2003), binary star formation, or a nonstandard nucleosynthesis 
(Oh et al. 2001) for Pop~III stars could solve this disagreement.

In order to constrain the mass range and explosion mechanism for Pop~III stars which
contribute to the chemical feedback on galactic scales, we have analyzed an extremely
metal poor protogalactic structure at high redshift by quasar absorption-line spectroscopy.  

Quasar absorption-line (QAL) systems are important objects to study 
the IGM at low and high redshifts. 
QAL systems sample both low and high density regions in the Universe and 
thus provide important information about structure formation
and early chemical evolution. Metal abundance measurements in QAL systems 
at high redshift are particularly important to learn about the
first generations of stars in the Universe, as these objects
should have left a characteristic signature in the abundance
pattern of chemically young systems.
Damped Lyman $\alpha$ (DLA) absorbers are QAL systems with large hydrogen 
column densities ($N$(H{\sc i})\,$\ge$\,$2\times 10^{20}\,$cm$^{-2}$), and 
most suitable for studies of the chemical evolution at high redshift (e.g., 
P\'{e}roux et al. 2003, Richter et al. 2005). Number statistics of DLA systems 
imply that these objects dominate the neutral gas content of the Universe at 
$z>1$ (Lanzetta et al. 1995; Wolfe et al. 1995; Rao \&  Turnshek 2000). While most of
the observable baryonic content of today's galaxies is concentrated in stars, in 
the past it must have been in the form of gas. The general agreement between the 
estimated baryonic mass density for DLA systems ($\Omega _{\mathrm {DLA}}$) at $z\approx2$ 
and the baryonic mass density in stars at $z=0$ makes DLA systems prime 
candidates to be the progenitors of present-day galaxies (Wolfe et al. 1995; 
Storrie-Lombardi \& Wolfe 2000). The most important absorption lines of atomic and 
molecular species in DLA systems at high redshift ($z\approx2$-$3$) fall into 
the optical band. Background quasars at these redshifts are usually faint 
($V$\,$\approx$\,17$^{\mathrm{m}}$-20$^{\mathrm{m}}$), however, so that
one needs 8-10\,m class telescopes to obtain high-resolution and high
signal-to-noise (S/N) absorption spectra that provide accurate
information on these objects.

The DLA system discussed in this paper is characterized by a very low overall 
metallicity. It exhibits an abundance pattern that points to an enrichment
of only one or at most a few stellar generations.
Thus, the UVES high-resolution data of the DLA system toward Q0913+072 
provide us with a unique insight into 
the early enrichment history of a proto-galactic structure 
at $z\approx2.6$.

\section{The Line of Sight toward Q0913+072}

An earlier analysis of the DLA system at $z_{\mathrm{abs}}=2.6183$
toward Q0913+072 ($V=17.1,$ $z_{\mathrm{em}}=2.785$) has been
presented by Ledoux et al.\,(1998). These authors have used the F/8 Cassegrain
focus of the ESO 3.6\,m telescope and the Nasmyth focus of the ESO 3.5\,m 
New Technology Telescope (NTT) at La Silla, Chile. A total neutral hydrogen 
column density of log$\,N$(H\,{\sc i})\,=\,20.2\,$\pm$0.1 was derived from 
a fit of the Ly\,$\alpha$ line. With a resolution of $R\sim13,000$, only a single component 
in the low-ionization absorption lines of O\,{\sc i}, C\,{\sc ii}, and 
Si\,{\sc ii} could be identified. An analysis of these lines suggested metal abundances
of $[\rm{O/H}]\approx -2.8$ and $[\rm{Si/H}]\approx-2.4$ together with a 
Doppler parameter of $b\approx7$\,km\,s$^{-1}$. At this resolution, however,
the true velocity structure in the gas is barely resolved. This leads to 
systematic uncertainties for the derived metallicities, since for 
several unresolved components the true $b$ values may be significantly lower,
and thus the column densities could be underestimated.
An analysis of this very interesting DLA system with higher-resolution data
holds the prospect of deriving more accurate abundances
for C, N, O, Si, and other elements and to investigate
the abundance pattern in this system in the context of the 
metal enrichment of protogalactic structures at high redshift.

\section{Metal Abundance Measurements}

The data used in this study were obtained
with the ESO Very Large Telescope (VLT) UV-Visual Echelle
Spectrograph (UVES) between January 17 and February 16, 2002, mostly during dark
time. The total integration time amounts to 6.8 hours (Dic1 + Dic2) while
the seeing-conditions were on average $1\arcsec$. Q0913+072 was observed
through a 1"-slit with two setups using dichroic beam splitters for the blue
and the red arm (Dic 1 with B390 and R580\,nm, and Dic 2 with B437 and R860\,nm,
respectively). This setup covers a wavelength range from
$\sim3250$-$10,200\,\AA$ with small gaps at $\sim5750$-$5840\,\AA$ and
$\sim8520$-$8680\,\AA$, respectively. 
The data have high spectral resolution ($R\sim45,000$) 
together with a high signal-to-noise ratio (S/N). The S/N typically                               
varies between $\sim30$-$70$ per pixel ($\mathrm{width}=1.9\,km\,s^{-1}$) over 
the wavelength range between $3500$-$6000\,\AA$. The CCD pixels were binned $2\times2$ and the
raw data were reduced using the UVES data-reduction pipeline implemented
in the ESO-MIDAS software package. These data are public and can be found in
the UVES database of ESO's Science Archive
Facility\footnote{http://archive.eso.org}.

The spectrum was analyzed using the FITLYMAN program (Fontana \& Ballester 
1995) implemented in the ESO-MIDAS software package. The routine uses a 
$\chi^2$-minimization algorithm for multi-component Voigt-profile 
fitting. Simultaneous line fitting of the high-resolution 
spectrum allows us to determine the column density, $N$, and the Doppler 
parameter, $b$, with high accuracy. The $b$ values of the DLA system toward 
Q0913+072  are assumed to be composed of a thermal component, $b_{th}$, and
a non-thermal component, $b_{non-th}$, in the way that
\begin{equation}
\label{eq1}
b=\sqrt{b_{\mathrm{th}}^2+b_{\mathrm{non-th}}^2}. 
\end{equation}
The non-thermal component may include processes like 
macroscopic turbulence, unresolved velocity components, and others. 
Note that in our case $b_{\mathrm{th}}^2 \ll b_{\mathrm{non-th}}^2$.

\begin{figure}[h!] 
  \centering
  \includegraphics[width=6.23cm, angle=0, clip] {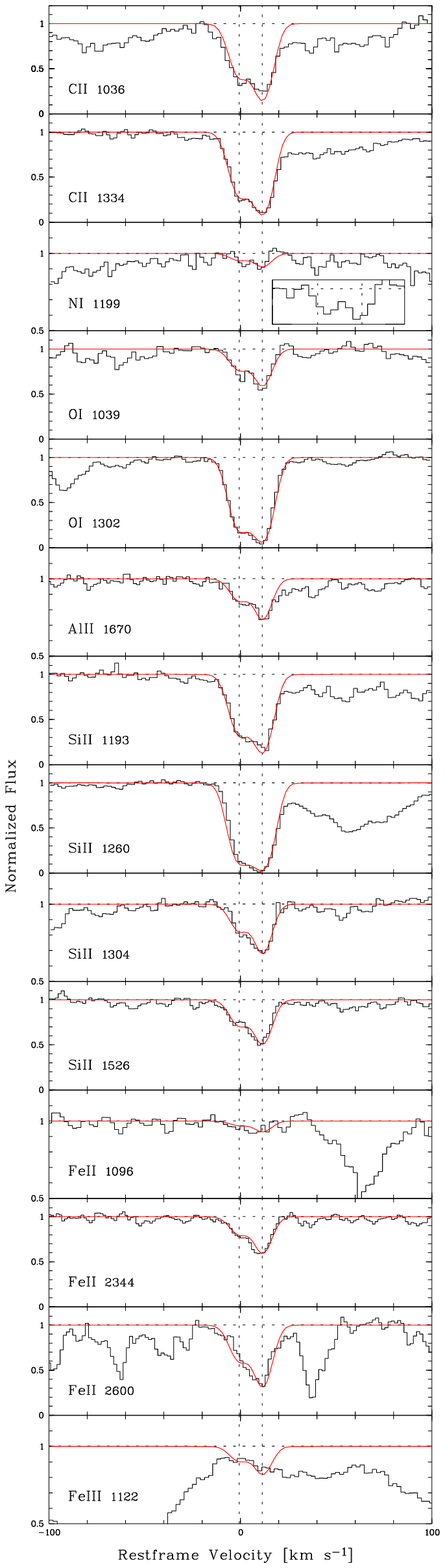}
  \caption[]{Absorption profiles of various ions are plotted against
             the restframe velocity. Two dominant absorption 
	     components are identified at $-1$ km\,s$^{-1}$ and
	     $+11$ km\,s$^{-1}$ (vertical dashed lines).}
  \label{lvplot}
\end{figure}

\begin{figure*}[] 
  \centering
  \includegraphics[width=15cm, angle=0, clip]{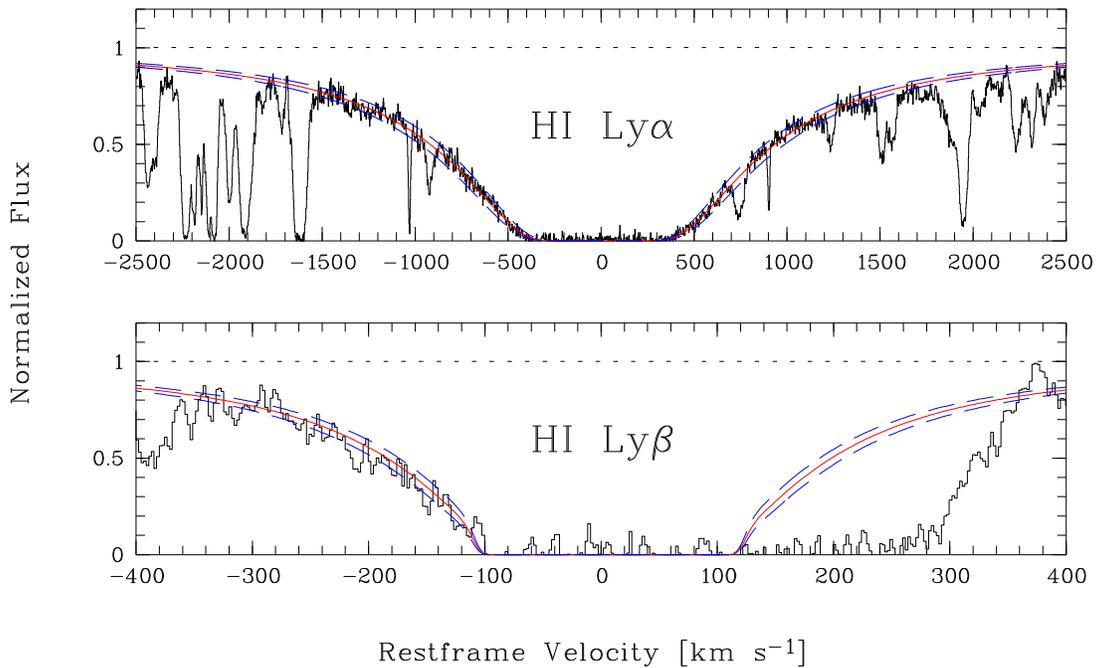}
  \caption[]{The Ly\,$\alpha$ and Ly\,$\beta$ absorption profiles 
             of the DLA system at $z_{\rm abs}=2.6183$ toward the quasar Q0913+072 are shown, 
	     plotted on a restframe velocity scale. The solid line shows
	     the optimum single-component fit with 
	     $\mathrm{log}\,N$(H\,{\sc i})\,=\,20.36,
	     the dashed lines indicate the $1\sigma$ error range of $\pm$0.05\,dex.}
  \label{hlines}
\end{figure*}

 
Due to the low overall metallicity of this system, metal absorption is detected
only in the strong lines of C\,{\sc ii}, N\,{\sc i}, O\,{\sc i}, Al\,{\sc ii}, 
Si\,{\sc ii}, and Fe\,{\sc ii}. No other low ion or other element could be identified.
Upper limits for elements like sulfur or zinc can be derived but they are of no 
further use. Fitting the detected low-ionization lines we can identify two 
dominant velocity components at $-$1\,km\,s$^{-1}$
and $+$11\,km\,s$^{-1}$ in the $z_{\mathrm{abs}}=2.6183$ restframe 
(Fig.\,\ref{lvplot}), as well as three satellite
components at $+$114, $+$152, and $+$181\,km\,s$^{-1}$, respectively.
For the study presented in this paper we will concentrate on the two dominant components.
Due to the large neutral hydrogen column density the H\,{\sc i} Ly\,$\alpha$ 
absorption spans several \AA, so that all sub-components are superposed in one big Lyman
trough (Fig.\,\ref{hlines}). Therefore, it is impossible to decompose the 
H{\sc i} absorption line profile into the various velocity subcomponents.
From a simultaneous single-component fit of the Ly\,$\alpha$
and Ly\,$\beta$ absorption we obtain a total
neutral hydrogen column density of log$\,N$(H\,{\sc i})\,=\,$20.36\pm0.05$. 
This value, together with the total column densities derived for the species 
listed above, is used to determine metal abundances
$[\mathrm{X/H}]=\mathrm{log}(N(\mathrm{X})/N(\mathrm{H}))-\mathrm{log}(N(\mathrm{X})/N(\mathrm{H}))_\odot$. 
These range from $-3.84$ (nitrogen) to $-2.47$ (oxygen),
implying that the overall metal abundance ($[\rm{M/H}]=[\rm{O/H}]=-2.47$ or $Z=0.0034$) of the 
gas is very low. All individual column densities and abundances in the 
$z=2.6183$ absorber are summarized in Table\,\ref{summary}. The errors 
in this paper represent the $1\,\sigma$ fitting errors given by the FITLYMAN program. 
Additional uncertainties arise from to the continuum placement (on the order of 0.1\,dex) and
the fixing of the line centers and $b$ values. Note that these additional error
sources are not included in the error estimates listed.

Our measurements confirm the previous abundance estimates from 
Ledoux et al.\,(1998). However, due to the higher resolution of the UVES data, the 
system is now resolved into two major and three satellite components 
for the low ions (instead of one), and five components for the
high ions (instead of two). The new detection of Al\,{\sc ii} and upper limits for 
Fe\,{\sc iii} and Ni\,{\sc ii} provide additional information about the 
chemical composition of the gas.
With $[\rm{M/H}]=-2.47$ ($\sim{^1\!/\!_{300}}$ solar) this absorber has one of the 
lowest metallicities ever measured in DLA systems. More precisely, it is among
the four lowest in the sample of 100 DLA systems from Prochaska et al. (2003) and 
the system with the lowest metallicity in the redshift range 2$\,<\,$z$\,<\,$3. Also, 
this DLA system is (to our knowledge) the only system for which an accurate carbon
abundance can be determined. Due to the low metallicity the strong C\,{\sc ii} 
absorption is not heavily saturated (or very mildly at most) in contrast to 
all other known DLA systems.

Note while the two-component structure is clearly seen in the 
N\,{\sc i}\,$\lambda$\,1199 line, the observed absorption profile slightly deviates
from the expected shape. The reason for this is unclear, but possibly related to noise
features that are present in this wavelength range (see Fig. \ref{lvplot}). The
detection significance exceeds $3\,\sigma$ and $4\,\sigma$ for the blue and the
red component, respectively. However, the main source of uncertainty for the column 
density derived from the N\,{\sc i}\,$\lambda$\,1199 line is related to the 
choice of the continuum level.

\begin{table} [h!]
   \setlength{\tabcolsep}{0.7mm}
  \caption[]{Summary of chemical abundances using the common notation
             $[\mathrm{X/H}]=\mathrm{log}(N(\mathrm{X})/N(\mathrm{H}))-\mathrm{log}
	     (N(\mathrm{X})/N(\mathrm{H}))_\odot$.
             The column densities represent the sum over the two
	     main absorption components at $-$1\,km\,s$^{-1}$
	     and $+$11\,km\,s$^{-1}$, as derived from simultaneous multi-component
	     Voigt-profile fits. The total neutral hydrogen column density
	     is $\mathrm{log}\,N$(H\,{\sc i})\,=\,$20.36\pm0.05$.	     	     
	     Oscillator strengths are taken from Morton (2003). Solar reference
	     abundances are listed in Table \ref{solar}.  
	     
	     }
  \centering
  {\small
  \begin{tabular}{l|cccc}
  \hline
  \hline

  Species        &Transition             &$\mathrm{log}\,N(\mathrm{X})\pm\sigma_{\mathrm{log}\,N}$  &$[\rm{X/H}]\pm\sigma_{\rm{[X/H]}}$  &    \\
                 &lines used             &                                     &                                    &    \\
  \hline
  H\,{\sc i}     &1025,1215              &20.36 $\pm$0.05                      &                                    &    \\           
  C\,{\sc ii}    &1036,1334              &14.05 $\pm$0.01                      &$-$2.83\,$\pm$0.05                  &    \\
  N\,{\sc i}     &1199                   &12.47 $\pm$0.10                      &$-$3.84\,$\pm$0.11                  &    \\
  O\,{\sc i}     &1039,1302              &14.58 $\pm$0.02                      &$-$2.47\,$\pm$0.05                  &    \\
  Al\,{\sc ii}   &1670                   &11.84 $\pm$0.01                      &$-$3.01\,$\pm$0.05                  &    \\
  Si\,{\sc ii}   &1304,1526              &13.35 $\pm$0.01                      &$-$2.57\,$\pm$0.05                  &    \\
  Fe\,{\sc ii}   &1096,2344,2600         &13.09 $\pm$0.01                      &$-$2.77\,$\pm$0.05                  &    \\
  Fe\,{\sc iii}  &1122                   &$\le$13.33                           &                                    &    \\
  Ni\,{\sc ii}   &1370                   &$\le$12.17                           &$\le-$2.44                          &    \\
  \hline
  \end{tabular}
  \label{summary}
  \noindent
  } 
\end{table}

\begin{table} [h!]
   \setlength{\tabcolsep}{0.7mm}
  \caption[]{Solar abundances as listed in Morton (2003), based on data from 
             Grevesse \& Sauval (2002), except for oxygen, for which we adopt the value 
	     given in Allende Prieto et al.\,(2001).} 
  \centering
  {\small
  \begin{tabular}{l|c}
  \hline
  \hline
   Element       &$\mathrm{log}N((\mathrm{X})/N(\mathrm{H}))_{\odot}+12.00$ \\ 
  \hline
  Carbon         & 8.52 \\ 
  Nitrogen       & 7.95 \\ 
  Oxygen         & 8.69 \\ 
  Aluminum       & 6.49 \\ 
  Silicon        & 7.56 \\ 
  Iron           & 7.50 \\ 
  Nickel         & 6.25 \\ 
   \hline
  \end{tabular}
  \label{solar}
  \noindent
  } 
\end{table}

\section{Results and Discussion}

\subsection{Metal abundance pattern}

The redshift of $z_{\mathrm{abs}}=2.6183$ of this system 
corresponds to a lookback time of
$11.1$ Gyr (i.e., to an age of the Universe of $2.5$ Gyr at that 
redshift)\footnote{using $H_{0}=71$\,km\,s$^{-1}$,
$\Omega _{\mathrm{m}}=0.27$, and $\Omega _{\mathrm{tot}}=1$} .
Given the very low overall abundance level and the observed abundance 
pattern, the chemical evolution of this system
is lagging behind the evolution of other DLA systems at similar redshifts.
The DLA system toward Q0913+072 can be compared with the nearby
dwarf galaxy \emph{I~Zwicky~18} (at a distance of $\sim 15$\,Mpc), which contains no 
stars older than 500~Myr (Izotov \& Thuan 2004). Either these kinds of galaxies have 
formed just recently, or they have been existing as protogalactic structures for 
several Gyr. An alternative explanation for the poor enrichment
we observe in the DLA system toward Q0913+072 would be that an initial 
violent starburst might have disrupted the integrity of the protogalactic 
structure and consequently would have prevented further star-formation activity. 
This would bring the nucleosynthesis processes to a halt and the chemical 
composition of the gas would reflect the enrichment pattern produced 
by the initial starburst.

\begin{figure}[h!] 
  \centering
 \includegraphics[width=8.5cm, angle=0, clip]{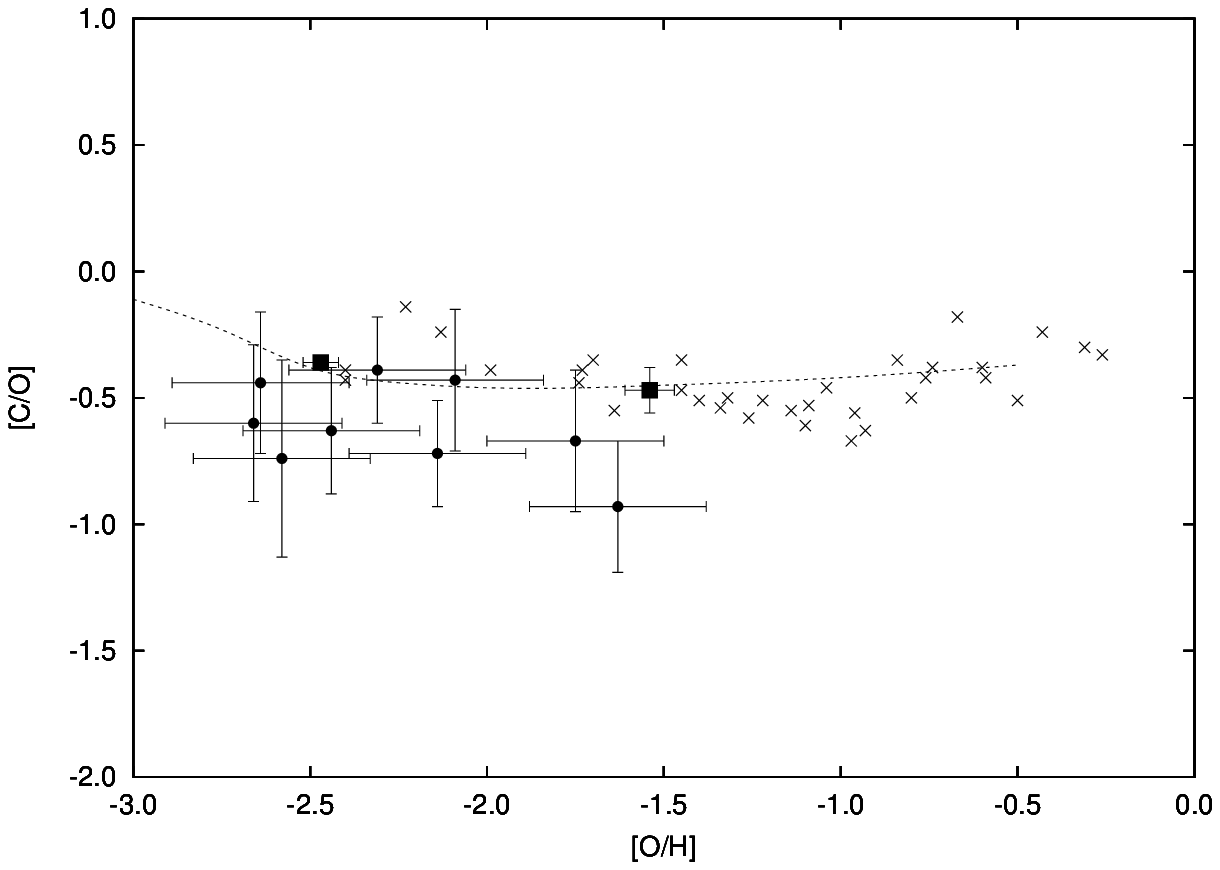} 
 \caption[]{We compare the only two existing carbon abundance measurements in DLA 
            systems (squares; from this study and from D'Odorico \& Molaro 2004) with 
	    extremely metal-poor halo stars (crosses) from 
	    Akerman et al. (2004), and (circles) from Spite et al. (2005). The dashed 
	    line shows predictions from Pop~III star yields from Chieffi \& Limongi (2002)
	    for the case of an IMF with $M_{\mathrm{ms}}\ge 10\,M_\odot$.
	}
  \label{cplot}
\end{figure}

In the following we focus on the abundances of individual elements
that could shed light on the chemical evolution history of this system.
\\
\\
\emph{Nitrogen} -- Comparing our results with a database of 66 DLA systems 
(Centuri\'{o}n et al. 2003, and Lanfranchi et al. 2003), the nitrogen abundance 
is among the lowest ever measured (see also Richter et. al 2005). There is 
general consensus that oxygen is almost entirely produced by (massive) SNe, 
the contribution from low- and intermediate-mass stars is practically irrelevant. 
Carbon, on the other hand, is produced in stars of nearly all masses (essentially 
by helium burning). In contrast to oxygen and carbon, however, the initial formation 
of nitrogen is still not well understood. If nitrogen is formed directly from 
helium it is called \emph{primary}. If the nitrogen production is dependent on 
pre-existing carbon seed nuclei, it is called \emph{secondary} and is obviously 
dependant on the star's initial metallicity. While all stars which reach the 
CNO cycle (nitrogen is formed at the expense of carbon and oxygen) produce 
secondary nitrogen, it cannot be decided at present whether primary nitrogen 
is produced mostly in massive stars (exploding as SNe) or in intermediate-mass 
stars ($\sim 4-8\,M_{\odot}$) during their AGB phase, or both 
(Spite et al. 2005). The role of
low-mass stars ($<4\,M_{\odot}$) for the nitrogen enrichment at high $z$
is not well understood yet, but needs to be explored
in future studies (see, e.g., Pettini et al.\,2002). Pettini et al.
suggest that a significant fraction of DLA systems with oxygen
abundances between $\sim{^1\!/\!_{10}}$ and $\sim{^1\!/\!_{100}}$ solar 
have not yet attained the full primary level of nitrogen 
enrichment (i.e., $\mathrm{log}\,[\rm{N/O}]<-1.5$) and may represent 
protogalactic structures that just recently have formed out of the IGM.
\\
For the $\alpha$-elements, the enrichment depends 
critically on the star formation rate (SFR). Different absorber systems 
have different star formation histories, which means that they can reach the same
amount of $\alpha$-enrichment at different epochs. However, the nitrogen 
production depends more critically on the lifetime of the progenitor stars rather 
than on the SFR. Intermediate-mass stars dominate the nitrogen production and 
will follow the massive stars ($M>8\,M_\odot$), which are the main production 
sites for carbon, with a typical lag time of $\sim$\,250\,Myr (Henry et al. 2000).
The two plateaus in a [N/Si,O] versus [N,O/H] plot for DLA systems
nicely reflect this bimodal nitrogen production scenario (see Prochaska et al. 2002). 
The observed range in the [N/O] ratio at low [O/H]              
then is a natural consequence of the delayed release of nitrogen
into the ISM relative to the oxygen produced by SNe.
In the DLA system towards Q0913+072 we measure $[\rm{N/O}]=-1.37\pm0.10$, i.e, 
primary nitrogen. This is what we would expect when observing a chemically 
young system that just has recently condensed out of the IGM. 
\\
Our value of $[\rm{N/O}]=-1.37\pm0.10$ does not favor a 
very small dispersion in low-N DLA systems, i.e., $[\rm{N}/\alpha]=-1.45\pm0.05$, as 
suggested earlier by Centuri\'{o}n et al. (2003). 
Matteucci \& Calura (2005) argue that Pop~III stars alone cannot be responsible for the 
abundance ratios in low-metallicity DLA systems. They conclude that the intermediate-mass 
Pop~II stars must have played an important role for
the early nitrogen enrichment in the Universe, even at redshift of $z=5$. 
However, the redshift alone is not a good indicator for the 
evolutionary state of a galaxy. The very low abundances measured in 
the DLA system toward Q0913+072 together with the very low nitrogen content, 
when compared with yields from explosive nucleosynthesis model calculations, 
indeed speak against a major contribution from intermediate mass
Pop~II stars to the nitrogen enrichment in this DLA system.
\\
\\
\emph{Carbon} -- Reliable carbon measurements in QAL systems are very sparse, since
in most cases the C\,{\sc ii} absorption in DLA systems, sub-DLA systems, and 
Lyman-Limit Systems (LLS)\footnote{$10^{17.2}$cm$^{-2}$\,$\le$\,$N$(H\,{\sc i})\,
$\le 10^{19}$cm$^{-2}$} is heavily saturated.
Levshakov et al. (2003) have measured C\,{\sc ii} absorption in a LLS 
at $z_{\mathrm{abs}}=2.917$\footnote{All
previous measurements of metal abundances in LLS lie in the
range of $[\rm{M/H}]\ge-2.4$ (Fan 1995; Songaila \& Cowie 1996; Levshakov 
et al. 2002a and 2003).}.
D'Odorico \& Molaro (2004) also reported a carbon measurement in a DLA system 
at $z_{\mathrm{abs}}=4.383$, but their C\,{\sc ii} and O\,{\sc i} lines 
presumably are mildly saturated. Although the results of the LLS depend on 
photo-ionization calculations, all cases seem to have $[\rm{C/Fe}]\sim0.0$ 
and $[\rm{C}/\alpha]<0$ in common. For the DLA system toward Q0913+072 we derive 
$[\rm{C/Fe}]=-0.06\pm0.01$, $[\rm{C/O}]=-0.36\pm0.02$, and $[\rm{C/Si}]=-0.26\pm0.01$,
thus very similar to what has been derived for the other systems.
In all four cases, carbon appears to be underabundant compared to the $\alpha$ elements.
Although LLSs and DLA systems are expected to trace different cosmological
objects (at least in general), the observed underabundance of carbon may provide 
information about the early metal enrichment in the Universe. Clearly, more 
carbon measurements in DLA systems and LLSs are desired to investigate this 
interesting behavior in detail.
\\
\\
\emph{Oxygen} -- Oxygen is the best element to infer the $\alpha$-element 
abundance in interstellar and intergalactic gas, as O\,{\sc i} and
H\,{\sc i} have similar ionization potentials and both elements are
coupled by a strong charge-exchange reaction.
In the most widely cited set of solar abundances from 
Grevesse \& Anders (1991), the solar oxygen abundance was reported to be 
$\mathrm{log}\,(N(\rm{O})$/$ N(\rm{H}))_\odot+12=8.93$. However, recent estimates
lead to significant revisions of the Sun's oxygen abundance. In this work we 
adopted the value of $\mathrm{log}\,(N(\rm{O})$/$N (\rm{H}))_\odot+12=8.69$ from 
Allende Prieto et al. (2001). Further, uncertainties in stellar nucleosynthesis 
inputs such as the $^{12}$C($\alpha$,$\gamma$)$^{16}$O reaction rate, 
convection, fall back and mass loss could alter 
these and previous results, which therefore have to be viewed with caution.
\\
\\
\emph{Dust Depletion} -- No trace of molecular hydrogen was found the DLA system toward 
Q0913+072. The logarithmic ratio of hydrogen nuclei in molecules to the total hydrogen nuclei is 
$\mathrm{log}\,f_{\mathrm{H_2}}=\mathrm{log}\,2N(\mathrm{H_2})-\mathrm{log}\,N(\mathrm{H})\le-6.9$, 
where $\mathrm{log}\,N(\mathrm{H_2})\le13.2$, i.e., the total upper limits of the rotational
ground states $J=0$ and $J=1$ in the Werner band of molecular hydrogen. Because of 
the lack of $\mathrm{H_2}$ (which predominantly forms on dust grains) and the extremely low 
overall metallicity we do not expect dust depletion to alter significantly the abundance 
pattern in this absorption system (Ledoux et al. 2003, Vladilo 2002). 
\\
\\
\emph{Photoionization Corrections} -- Given the large neutral hydrogen
column density in this absorber, it is not expected that photoionization
has any significance on the abundances listed in Table\,\ref{summary}. This is supported
by our photoionization model for a DLA system at $z=2.5$ based on CLOUDY (Ferland et al. 1998).
In the case that photoionization should nonetheless be relevant, i.e., in view of
the observed two-component structure, the fact that the ionization potential (IP) 
of H\,{\sc i} is lower than the IPs 
of C\,{\sc ii}, N\,{\sc i}, Al\,{\sc ii}, Si\,{\sc ii}, and Fe\,{\sc ii}, means
that correcting for ionization effects would further decrease these abundances.  
\\
\\
\emph{Comparison with Metal-Poor Stars} -- It is legitimate to compare very metal-poor 
DLA systems to very metal-deficient stars which are believed to retain information 
of a preceding single SN event or at most a few (McWilliam et al. 1995). 
Comparing the abundances from this DLA system with a 
sample of 9 (unmixed) extremely metal-poor (EMP) Galactic halo giants 
($-4.0\le[\rm{Fe/H}]\le-2.0$) from Spite et al. (2005), we find a very good 
agreement for carbon, a reasonable agreement for oxygen and iron (typically $\pm0.2$\,dex), 
but clearly higher values for nitrogen (in average of $\sim$\,1\,dex) in the EMP stars. 
In Fig.\,\ref{cplot}, we include a sample of 34 F and G dwarf and 
subgiant stars belonging to the halo population 
($-3.2\le[\rm{Fe/H}]\le-0.7$) from Akerman et al. (2004). 
Although their mean values for [C/H] and [O/H] are higher by more than 1\,dex, 
their lowest values for carbon and oxygen are very similar to our observations.

An interesting aspect is to compare the two existing carbon measurements in 
DLA systems (this paper and D'Odorico \& Molaro 2004) with these halo stars 
(Fig.\,\ref{cplot}). We find a reasonable agreement for the carbon evolution 
in the DLA systems and the halo stars. When comparing with predictions from yields 
from Pop~III stars (Chieffi \& Limongi 2002) for the case of an IMF with 
$M_{\mathrm{ms}}\ge 10\,M_\odot$ (Fig.\,\ref{cplot}, dashed line), we notice 
that the carbon abundances in the two DLA systems match these theoretical 
predictions extremely well.

Ultra-metal-poor giants with high [O/Fe] ratios, such as presented by Israelian et 
al. (2004), seem to predict a new type of SNe (Aoki et al. 2002) where most of 
the matter was absorbed by the iron core. Smaller [O/Fe] ratios possibly indicate that 
a significant fraction of iron was incorporated in the SN ejecta, leading to a smaller 
mass cut and thus less compact remnants, possibly neutron stars. Stars with very low 
metallicities and DLA systems obviously are extremely important objects to derive 
constraints on Pop~III stars. The nitrogen production mechanism clearly plays a key 
role but yet requires more detailed work.


\subsection{Additional Constraints from Numerical Models}

The DLA system toward Q0913+072 is a chemically very young DLA system,
enriched by one or at most a few stellar generations that have left their 
typical abundance signature in the gas. We now want to 
compare the abundance pattern of this system with predictions 
from recent model calculations to learn more about the origin 
of this absorber.
\\
\\
\emph{SNe~Ia vs. SNe~II} -- 
The absence of a clear-cut [$\alpha$/Fe] ratio leaves the possibility 
open that SNe~Ia might have contributed to the metal enrichment of the 
observed DLA system, especially to the iron-peak elements. On the 
other hand, there are arguments against any significant 
contribution from SNe~Ia: (1) Based on results from the \emph{Hubble Higher z 
Supernova Search}, Strolger et al. (2004) constrain the delay times for SNe~Ia, 
i.e., the time from the formation of the progenitor stars to the explosion, in 
the range of 2-4 Gyr, where the lower barrier of $2$ Gyr distinctly exceeds 
the age of the DLA system toward Q0913+072, recalling that the age of the Universe
at $z_{\mathrm{abs}}=2.6183$ was only 2.5 Gyr. (2) Metallicity effects influence the 
SNe~Ia rate. This kind of low-metallicity inhibition of SNe~Ia could occur 
if the iron abundance of the accreted matter from the companion is as low 
as $[\rm{Fe/H}]\lesssim-1$ (Kobayashi et al. 1998 and references therein). The companion 
star in a single-degenerated scenario is typically a red-giant (RG) with an initial mass of 
$M_{RG}\sim1\,M_\odot$ or a near main-sequence star with an initial mass 
of $M_{ms}\sim2$-$3\,M_\odot$. Both stars are not massive enough to ignite silicon burning. 
Consequently, an iron abundance, as low as measured in our case, must inhibit 
the occurrence of SNe~Ia.
The observed low [O/Fe] ratios (more generally: the [$\alpha$/Fe] ratios) thus should 
reflect the enrichment pattern of the first generations of stars that have
enriched this protogalaxy, most likely by SNe~II and probably 
HNe as well. We observe $[\rm{O/Fe}]=+0.30\pm0.02$ and $[\rm{O/Si}]=+0.10\pm0.02$ 
which is consistent with the yield ratios of SNe~II from massive stars from 
Kobayashi et al. (2005). Furthermore, the ratios of 
$[\rm{Si/H}]=-2.57\pm0.05$ and $[\rm{Fe/H}]=-2.77\pm0.05$ are roughly consistent 
with numerical investigations from Kawata et al. (2001), which predict that a 
typical SN II is producing more silicon ($\sim5:1$) and slightly less iron 
($\sim0.8:1$) than a typical SN Ia.
\\
\\
\emph{Very Massive Stars (VMS)} -- VMS with $M_{\mathrm{ms}}\approx130$-$300\,M_\odot$, 
exploding as PISNe, show a tendency to produce 
too much silicon, but not enough carbon and oxygen when compared to this DLA system.
The observed $[\rm{Si/C}]=+0.26\pm0.01$ or $[\rm{C/O}]=-0.36\pm0.02$, and
$[\rm{Si/Al}]=+0.44\pm0.01$ ratios imply an IMF in 
which the first stars cannot have been very massive (Tumlinson et al. 2004).  
\\
\\
\emph{Yields from explosive nucleosynthesis models} -- The metallicity of the 
DLA system we probe is only slightly higher than what cosmological models
predict for the IGM at this redshift. We speculate that this primeval metal 
enrichment we observe originates from only one single star generation, i.e.,
from Pop~III stars. If there was only one enrichment cycle, or at most
a few, then the ISM will carry the imprint from these stars and their typical 
enrichment pattern. As mentioned earlier, only stars in the range of 
$M_{\mathrm{ms}}\sim 10$-$50\,M_\odot$ and $M_{\mathrm{ms}}\sim 140$-$260\,M_\odot$
will contribute to the enrichment on galactic scales. In other words, 
the questions we ask is: were the first stars massive ($\sim 10$-$50\,M_\odot$) 
or super-massive ($\sim 140$-$260\,M_\odot$), i.e., were they core-collapse S/HNe 
or PISNe? 

We make use of the latest yield calculations from explosive nucleosynthesis 
models from Kobayashi et al. (2005) for the case of "ordinary" 
(core-collapse) SNe, the typical SNe~II, with an explosion energy of 
$E=1\times10^{\mathrm{51}}$\,erg $=E_{51}$, hyper-energetic 
(i.e., $E>E_{51}$) core-collapse HNe, and PISNe. Two mass-energy relations were 
set: a constant $E_{51}=1$ for SNe with 13, 15, 18, 20, 25, 30, and 40\,$M_\odot$, and 
$E_{51}=$10, 10, 20, and 30 for HNe with 20, 25, 30, and 40\,$M_\odot$, respectively. 
The  yields for PISNe with 150, 170, 200, and 270\,$M_\odot$ were taken from 
Umeda \& Nomoto (2002). 

\begin{figure}[h!] 
  \centering
 \includegraphics[width=8.5cm, angle=0, clip]{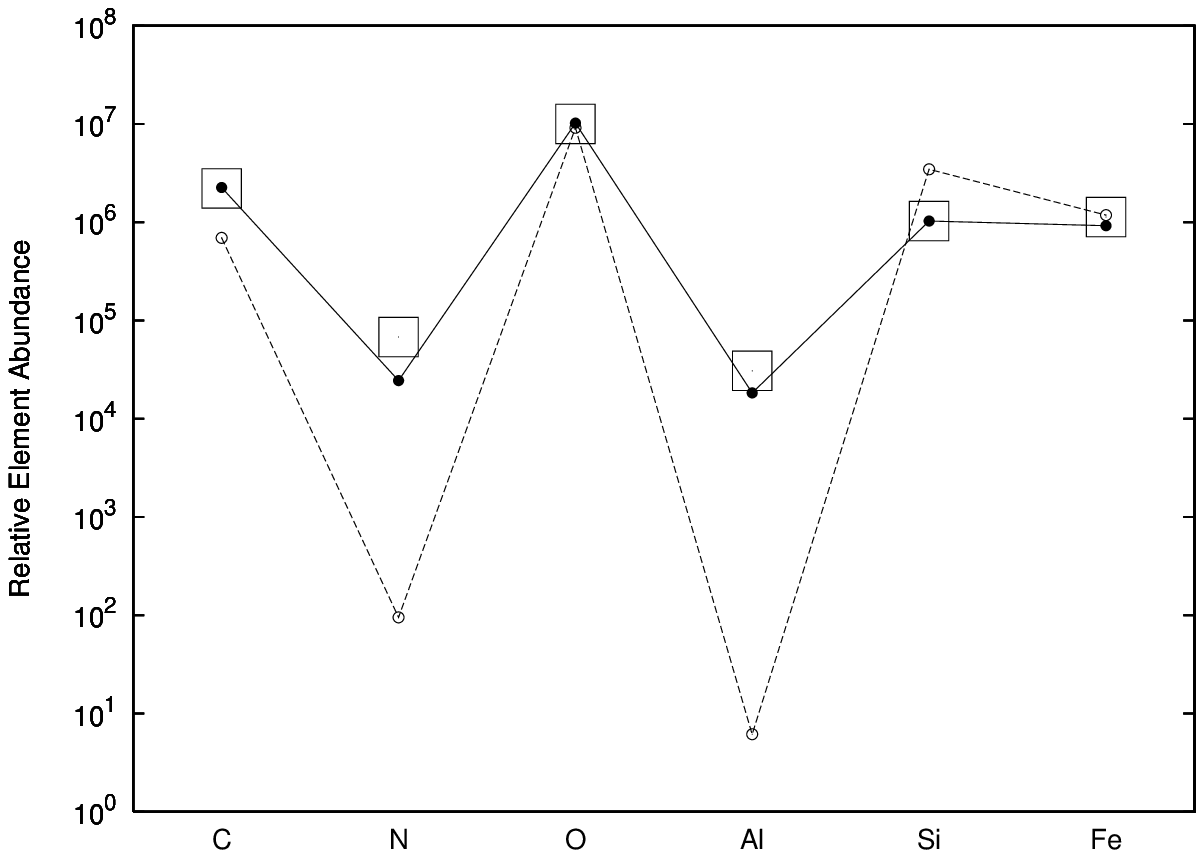} 
 \caption[]{Comparison of relative element abundances from the DLA system toward Q0913+072
            and model calculations for explosive nucleosynthesis. The open boxes show 
            the measured abundances in the DLA system toward Q0913+072 where the box size
	    corresponds to an assumed \emph{total} error of $\pm0.2$\,dex. Yields from 
	    numerical model calculations of a core-collapse SN with $15\,M_\odot$ 
	    and $1\,E_{51}$ are indicated with filled circles and continuous lines, and yields 
	    of a PISN with $200\,M_\odot$ are shown as open circles with dashed lines. 
	    The reader should note that \emph{relative} abundances are compared. For this
	    the sum of the abundances for C, N, O, Al, Si, and Fe of each set was
	    normalized to unity and then shifted arbitrarily ($\rm{oxygen_{\,DLA}}=10^7$)
	    to the scale shown on the y axis to indicate the full range of the 
	    measured and predicted element abundances. 
	}
  \label{DLAvsyields}
\end{figure}

The metal abundance pattern of this DLA system was then compared with these yields 
for metallicities of $Z=0$, $0.0001$, and $0.004$. We assumed a simple power-law IMF, 
i.e., $\xi(M_\star)=(M_\star/M_\odot)^\gamma$, 
varying $\gamma$ from $-5$ to $+2$ ($\gamma=-2.35$ corresponds to the Salpeter IMF).
We found that there is no significant difference between a scenario of SNe combined
with HNe and a scenario of SNe only. On the other hand, $Z$ and $\gamma$ are very
sensible indicators. We find that zero-metallicity stars produce
the most accurate abundance pattern in the case of a steep IMF
($\gamma\lesssim\gamma_{\mathrm{Salpeter}}$). This and the fact that yields from
PISNe do not match the abundance pattern observed in the DLA system toward Q0913+072 (see 
Fig.\,\ref{DLAvsyields}) excludes any significant contribution 
of PISNe, yet puts a question mark on their existence. Considering the case of an  
initially enriched ISM, we compared as well with yields from SNe and HNe with progenitors 
stars having $Z=0.0001$ and $Z=0.004$. For these second-generation stars (exploding as S/HNe), 
the abundance pattern is generally reproduced with less accuracy but we notice that a top-heavy 
IMF partially can reduce these inaccuracies.   

The reader should not take this analysis as a proof of the inexistence of PISNe but
rather as a strong indication that PISNe events must have been - at most - very rare. 
We also have to recall that the explosion energy, as well as the mass cut (i.e., the limiting radius 
between the remnant and the ejecta) for the core-collapse scenario are still two important 
parameters which are not very well understood, yet could lead to noticeable errors. 

Kobayashi et al. (2005) are using basically the same code as Umeda \& Nomoto (2001, 2002)
but with a different treatment of the mixing-fallback mechanism in order to explain 
the chemical evolution of the solar neighborhood. Umeda \& Nomoto, on the other hand, 
focus on iron peak elements, and, in particular, try to reproduce the large $[\rm{Zn/Fe}]$ 
observed in EMP stars. Using  the yields from Umeda \& Nomoto we find that the IMF should 
peak around stars with $25\,M_\odot$ and $10\,E_{51}$ in order to reproduce the abundance 
pattern we observe in this DLA system. When using the results from Kobayashi et al. we find 
that the IMF should peak around stars with typically $15\,M_\odot$ and $1\,E_{51}$.

\vspace{4mm}
\emph{Galactic and cosmic chemical evolution} -- In a recent paper Daigne et al. (2004) 
have modeled the chemical evolution of condensed cosmic structures (galaxies)
and the IGM as a function of redshift including the
process of reionization. They consider various different SFRs and IMFs for the 
progenitor stars that drive the metal enrichment in the early Universe.
Following their calculations, a bimodal (or top-heavy) IMF with a moderate 
mass range of $40$-$100\,M_\odot$ yields both, the required number of ionizing
photons to reionize the Universe at $z=17$, and the correct chemical composition   
of nucleosynthesis products of these stars to match the observations of
metal abundances in metal-poor halos stars and in the IGM.
When comparing the mass fractions of C, N, O, Si, and Fe presented in their
models with those derived for the DLA system toward Q0913+072 we find that
the abundances of these elements in the DLA system are slightly ($<1$\,dex,
typically) above the values predicted for the IGM at $z=2.6$, but 
substantially lower than what is expected for the ISM in galaxies 
at that redshift. This implies that the DLA system toward Q0913+072
represents a protogalactic structure that has just recently  
formed, so that local star formation activity 
in this system had not enough time to significantly enhance the abundance
level above that of the surrounding IGM. 

\section{Summary and Conclusions}

The DLA system at $z=2.6183$ toward Q0913+072 
is characterized by a very low overall metal abundance
($[\rm{M/H}]=[\rm{O/H}]=-2.47$ or $Z=0.0034$), a pronounced deficiency of nitrogen
($[\rm{N/H}]=-3.84$), and a mild underabundance of carbon
($[\rm{C/H}]=-2.83$). These values mark this system as the most metal-poor
DLA system observed at $z<3$. We further report 
this absorber to be the first DLA system at present date with a reliable carbon 
measurement. The low nitrogen abundance implies that this system has not yet attained 
the full level of primary nitrogen enrichment. This, and the
fact that the overall abundance level is only slightly above that of the 
IGM at that redshift, point toward a protogalaxy that has just
recently condensed out of the IGM.

The comparison of the abundances in this DLA system with  
a sample of extremely metal-poor halo giants shows 
a very good agreement for carbon, and a reasonable agreement 
for oxygen and iron. Predictions from yields from 
Pop~III stars (Chieffi \& Limongi 2002) for the case of an IMF 
with $M_{\mathrm{ms}}\ge 10\,M_\odot$ match the carbon 
measurements of the two available DLA systems extremely well.  

Stars with main-sequence masses below $\sim10\,M_\odot$ 
do not essentially contribute to the metal budget in the gas on galactic scales, 
and primordial stars in the range of $M_{\mathrm{ms}}\sim 50$-$140\,M_\odot$  
and above $\sim260\,M_\odot$ are believed to collapse without
being able to eject any enriched gas. The only stars with a 
chemical feedback must therefore be stars with $M_{\mathrm{ms}}\sim 10$-$50\,M_\odot$, 
exploding as core-collapse S/HNe, or stars with progenitors in the mass 
range $\sim140$-$260\,M_\odot$, exploding as PISNe. 
Scenarios of progenitor stars with
a non-zero metallicity, a top-heavy IMF, other explosion mechanism like
SNe~Ia or PISNe, or a combination of the foregoing, cannot satisfactorily
explain the observed chemical abundance pattern.
Comparing the metal abundance pattern of the DLA system at $z_{\mathrm{abs}}=2.6183$ 
toward Q0913+072 with yields from explosive nucleosynthesis model calculations, we 
conclude that the most likely scenario for the observed abundance pattern 
is massive (but not super-massive) Pop~III stars with $M_{\mathrm{ms}}\approx 10$-$50\,M_\odot$, 
exploding as core-collapse S/HNe.

\smallskip
\smallskip
\smallskip
\emph{Acknowledgements.} 

The authors are grateful to the referee Giovanni Vladilo for his
helpful report and insightful comments.
We also wish to thank ESO for the use 
of the UVES spectrum of Q0913+072 (ProgId 68.B-0115(A), 
01 Oct 2001, VLT-Kueyen, Paranal, Chile). 
PE and PR acknowledge financial support by
the German \emph{Deutsche Forschungsgemeinschaft}, DFG, through
Emmy-Noether grant Ri 1124/3-1.

\end{document}